\documentclass[twocolumn,prl,aps,superscriptaddress,showpacs]{revtex4}
\usepackage{xspace,amsmath,amsfonts,amsthm,amssymb,amsbsy,graphicx,color}

\newcommand{\beq}{\begin{equation}} 
\newcommand{\eeq}{\end{equation}}
\newcommand{\bqa}{\begin{eqnarray}} 
\newcommand{\eqa}{\end{eqnarray}}
\newcommand{\nn}{\nonumber} 
\newcommand{\nl}[1]{\nn \\ && {#1}\,}
\newcommand{\erf}[1]{Eq.~(\ref{#1})}

\newcommand{\bra}[1]{\left\langle{#1}\right |} 
\newcommand{\ket}[1]{\left |{#1}\right\rangle}
\newcommand{\ip}[2]{\left\langle{#1}|{#2}\right\rangle}
\newcommand{\op}[2]{\ket{#1}\bra{#2}}
\newcommand{\sch}{Schr\"odinger }

\newcommand{\sq}[1]{\left[ {#1} \right]}
\newcommand{\cu}[1]{\left\{ {#1} \right\}}
\newcommand{\ro}[1]{\left( {#1} \right)}

\newcommand{\tr}[1]{{\rm Tr}\left[ {#1} \right]}
\newcommand{\inty}{\int_{-\infty}^\infty \!}
\newcommand{\Dio}{Di\'osi}
\renewcommand{\section}[1]{{\em #1}.---}

\begin{document}

\title{Pure-state quantum trajectories for general non-Markovian systems do not exist }

\author{Howard M. Wiseman}
\affiliation{Centre for Quantum Dynamics, School of Science, Griffith University, Nathan 4111, Australia} 

\author{J. M. Gambetta}
\affiliation{Institute for Quantum Computing and Department of Physics and Astronomy, University of Waterloo, Waterloo, Ontario, Canada N2L 3G1}

\begin{abstract}
Since the first derivation of non-Markovian stochastic \sch\ equations, their interpretation has been contentious. In a recent Letter [Phys. Rev. Lett. {\bf 100}, 080401 (2008)], Di\'osi claimed to prove that they generate ``true single system trajectories [conditioned on] continuous measurement''. In this Letter we show that his proof is fundamentally flawed: the solution to his non-Markovian stochastic \sch\ equation at any particular time can be interpreted as a conditioned state, but joining up these solutions as a trajectory creates a fiction. 
\end{abstract}

\maketitle 

It is well recognized that the continuous measurement of an open quantum system $S$ with Markovian dynamics can be described by a stochastic \sch\ equation (SSE). The pure-state solution to such an equation over some time interval, a ``quantum trajectory'' \cite{Car93b}, can be interpreted as the state of $S$ evolving while its environment is under continuous observation (monitoring). This fact is of great importance for designing and experimentally implementing feedback control on open quantum systems \cite{adaptph,QED,spin-sqz}. If this interpretation could also be applied to {\em non-Markovian} SSEs \cite{StrDioGis,GamWis02}, then this would be very significant for quantum technologies, especially  in condensed matter environments, which are typically non-Markovian \cite{BrePet02}.


Previously we have argued that non-Markovian SSEs (NMSSEs) {\em cannot} be interpreted in this way \cite{GamWis02,GamWis03}. The solution at any particular time can be interpreted as the system state conditioned upon some measurement of the environment \cite{GamWis02}. But connecting up those solutions to make a trajectory is a fiction akin to trajectories in Bohmian mechanics \cite{GamWis03}. 
Restricting to standard quantum mechanics, 
the basic problem is that for the state of $S$ to remain pure, the bath field must be continuously observed to disentangle it from the system. For Markovian dynamics, this is not a problem, because the moving field interacts with $S$ and, having interacted, moves on. But for non-Markovian  dynamics, the field comes back and interacts again with $S$. Thus monitoring the field will feed disturbance back into the system, changing the {\em average} evolution of the state of $S$. That is contrary to the derivation of the NMSSE, which is constructed so as to reproduce on average the no-measurement evolution of $S$.

Recently, \Dio\ rederived one form of NMSSE from a different starting point, and claimed that, contrary to the conclusions of Ref.~\cite{GamWis03}, this allows an interpretation of the solutions as ``true single system trajectories [conditioned on] continuous measurement'' \cite{Dio08}. Here we show by general argument, and an explicit calculation, that this claim is incorrect, and that the reformulation does not alter our earlier conclusion.

\section{The non-Markovian system} \Dio\ considers a bath comprising an infinite sequence of von Neumann apparatuses $A_n$, each described by position and momentum operators $\hat x_n$, $\hat p_n$, $n\in\cu{1,2,\ldots \infty}$. (For clarity, we are using slightly different notation from Ref.~\cite{Dio08}.) The system interacts with the bath via the coupling Hamiltonian 
\beq \label{defV}
\hat V = \sum_n \delta(t-\tau_n) \hat X \hat p_n, \;\; \tau_n=\epsilon n,
\eeq
where $\hat X$ is an Hermitian system operator. Here the explicit time-dependence plays the role of the free propagation of a bath field. This would seem to be a recipe for generating Markovian evolution, since $S$ interacts only once with each $A_n$, which thus plays a role analogous to a small segment of a Markovian bath field. The novelty of \Dio's approach is to generate non-Markovian evolution by having the $\cu{A_k}_{k=1}^\infty$ prepared in an entangled state $\ket{\phi_0}$. In the position representation it is given by
\beq \label{phi0}
\ip{ \cu{x_k}_{k=1}^\infty}{\phi_0} \propto \exp\left[{ - \epsilon^2\sum_{l,m} x_l x_m \alpha(\tau_l-\tau_m) }\right].
\eeq
The continuum-time limit is $\epsilon\to0$, where the system is subjected to infinitely frequent, but infinitesimally strong, interactions with the apparatuses. In this limit, $\alpha(t)$ plays the role of the correlation function for the bath. 
 It is a real and symmetric function \cite{GamWis02,Bassi:2002a}, and equals $g^2\delta(t)$ in the Markovian case. Assuming the system is initially in a pure state also, the Hamiltonian (\ref{defV})  produces an entangled system--bath state $\ket{\Psi(\tau_n^+)}$ immediately after the $n{\mathrm{th}}$ interaction.

\Dio\ first considers the case where, immediately after each time $\tau_n$, the observable $\hat x_n$ is measured, yielding result $x_n$. This gives an unnormalized state for the conditioned quantum  system, 
$\tilde \rho(\tau_n^+;\cu{x_l}_{l=1}^{n})$, given by 
\beq 
  {\rm Tr}_{\cu {A_m}_{m=n+1}^{\infty}}\left[\ip{\cu{x_l}_{l=1}^{n}}{\Psi(\tau_n^+)}\ip{\Psi(\tau_n^+)}{\cu{x_l}_{l=1}^{n}}\right],
\eeq 
with  $\tr{\tilde \rho(\tau_n^+;\cu{x_l}_{l=1}^{n})}$ being the probability for the record $\cu{x_l}_{l=1}^{n}$. In the limit $\epsilon \to 0$, this state (if appropriately scaled) will have a continuous but stochastic evolution through time. The measurement of observable $\hat x_n$ does not disturb the future evolution of $S$ because $A_n$ never interacts with $S$ again. Thus, there is no difficulty with interpreting this stochastic evolution as the trajectory of an individual system, with the average state at time $t$ 
\beq
\rho(t) = \inty dx_0 \cdots \inty dx_n\ \tilde \rho(t;\cu{x_l}_{l=1}^{\lfloor t/\epsilon \rfloor})
\eeq
  being identical with that obtained simply by tracing over the bath (the apparatuses),
\beq
\rho(t) =   {\rm Tr}_{\cu {A_k}_{k=1}^{\infty}}\left[\ket{\Psi(t)}\bra{\Psi(t)}\right].
\eeq

It is obvious, however, that $\tilde \rho(t;\cu{x_l}_{l=1}^{\lfloor t/\epsilon \rfloor})$ is {\em not} the solution of a SSE, for the simple reason that the state is mixed, not pure, even if it begins pure \footnote{It is not clear whether this mixed state trajectory is the solution of a well-defined non-Markovian stochastic master equation.}.  The mixedness arises because the interaction of $S$ with $A_n$ entangles $S$ with $A_m$ for $m > n$, because initially $A_n$ and $A_m$ are entangled. That is, the system becomes entangled with apparatuses that are not yet measured. A mixed conditional equation state is not unexpected for non-Markovian systems. It has previously been shown in Refs. \cite{Imamoglu:1994a} and \cite{Breuer:2004b} that it is possible to derive a mixed state quantum trajectory equation that reproduces the non-Markovian evolution on average by adding to $S$ a fictitious system $F$, with the latter coupled to a monitored (Markovian) bath. A mixed state for $S$ arises when the partial trace over $F$ is performed. See Ref.~\cite{GamWis02b} for a comparison of this method with that of the NMSSE.

\section{The non-Markovian SSE and its interpretation} 
The only way to obtain a pure state for $S$ at time $t$ is by  measuring all the apparatuses with which the system is entangled. Specifically, \Dio\ shows that it is necessary to measure the set of bath observables $\cu{\hat z(s):s\in[0,t]}$, where 
$\hat z(s)$ is the ``retarded observable'' \cite{Dio08}
\beq \label{defz}
\hat z(s) = 2\epsilon \sum_{k=1}^{\infty } \hat x_k \alpha(s-\tau_k) .  
\eeq 
This is of course a different observable at different times $s$. The state conditioned on the result $Z_t \equiv \cu{z(s):s\in[0,t]}$ of this measurement at time $t$ is a {\em functional} 
of $z(s)$ for $0\leq s \leq t$, which we will write as $\ket{\bar \psi_t[Z_t]}$.  \Dio\ shows that this state is pure, and that it is the solution of the NMSSE  
\beq
\frac{d\ket{\bar \psi_t[Z_t]}}{dt} = \hat X_t\left(z(t) {-} 2 \int_0^t \alpha(t-s) 
\frac{\delta}{\delta z(s)} ds\right)\ket{\bar \psi_t[Z_t]}. \label{NMSSE}
\eeq
Here, \Dio\ is working in the interaction picture with respect to the system Hamiltonian $\hat H$; hence, the time dependence of $\hat X_t \equiv e^{i\hat H t}\hat X e^{-i\hat H t}$.
Equation (\ref{NMSSE}) was first derived in Refs.~\cite{GamWis02,Bassi:2002a}, but is very similar to that derived earlier in  Refs.~\cite {StrDioGis}.  The ensemble average of solutions of this NMSSE reproduces the reduced state of the system:
\beq
\rho(t) = {\rm E} \sq{\op{\bar \psi_t[Z_t]}{\bar \psi_t[Z_t]}} .
\eeq
Here in taking the expectation value, $z(t)$ must be treated as a Gaussian noise process with correlation function ${\rm E}[z(t)z(s)] = \alpha(t-s)$, as appropriate for $\ket{\phi_0}$. This convention is indicated by the notation $\bar\psi$ (as oppiosed to $\tilde\psi$) for the state. 

The contentious issue is not whether the solution  $\ket{\bar \psi_t[Z_t]}$ has an interpretation in standard quantum mechanics. As just explained, this state is the conditioned state of $S$ at time $t$ if an {\em all-at-one measurement} of the set of bath observables $\cu{\hat z(s):s\in[0,t]}$ were made at that time, yielding the result $Z_t$.  The contentious issue is: can the {\em family} of states $\ket{\bar \psi_t[Z_t]}$ for $0\leq t \leq \infty$ be interpreted as a trajectory for the state of a single system, conditioned on monitoring of its bath. \Dio\ claims that it can be so interpreted, and that the required monitoring is simply to measure $\hat z(\tau_0)$ at time $\tau_0^+$, $\hat z(\tau_1)$ at time $\tau_1^+$ and so on. At first sight this {\em monitoring} may seem equivalent to the all-at-once measurement described above. But in fact it is not, as we will now explain. 

A measurement of $\hat z(t)$ at time $t^+$ involves measuring apparatuses that have not yet interacted with $S$. This is necessarily so  because the symmetry of $\alpha(\tau)$ means that $\hat z(t)$ contains contributions from $\hat x_m$ for some $\tau_m>t$ (except for the Markovian case of course). Consequently, $\hat z(t)$ does not commute with $\hat p_m$ for some $\tau_m>t$, and the measurement will therefore disturb these momentum observables. But these are precisely the observables that will couple to the system via (\ref{defV}), and thereby disturb it. Thus, as soon as the first measurement is performed, of $\hat z(\tau_0)$ at time $\tau_0$, $S$ ceases to obey the NMSSE. Whatever stochastic evolution it does undergo, it will not reproduce the reduced state of the unmeasured system $\rho(t)$. 



It might be thought that it would be possible to avoid this alteration of the future evolution of the system by repreparing the apparatuses $A_m$ for $\tau_m>t$ in their pre-measurement states. However, this is not possible; before the measurement, these $A_m$ were {\em entangled} with the system $S$ and the other apparatuses. The correlation of these $A_m$ with $S$ is why the system state $\tilde \rho(\tau_n;\cu{x_l}_{l=1}^{n})$,   conditioned on measuring the apparatuses after they have interacted with the system, is {\em mixed}. The evolution of this state over time is the only true quantum trajectory for a single system, and its mixedness is an inevitable consequence of the non-Markovian dynamics. In fact, we now show by explicit calculation that the monitoring \Dio\ suggests does not even produce pure conditioned states of $S$ --- it also leads to mixed states. 

\section{A simple example} 
We consider the case where the bath consists of two apparatuses and $\epsilon=1$.
Thus  there are just three relevant times, $\tau_0=0$ (the initial time), $\tau_1^+=1$ (just after the interaction with $A_1$) and $\tau_2^+=2$ (just after the interaction with $A_2$). Without loss of generality, we can write the initial Gaussian entangled state of the bath, analogous to \erf{phi0}, as 
\beq
\phi_0(x_1,x_2)=c\exp[-(x_1^2+x_2^2 + 2 ax_1x_2)],
\eeq
where $c^2 ={2 \sqrt{1-a^2}}/\pi$. Here $0\leq a<1$ parametrizes the initial entanglent between the apparatuses.  The analogue of \erf{defz} defines two operators, 
\begin{equation}
	\begin{split}
		\hat z_1 = 2(\hat x_1+ a\hat x_2) ,  ~\hat z_2 = 2(\hat x_2+ a\hat x_1)  .
	\end{split} \label{defdisz}
\end{equation}

Let us consider the unconditioned evolution of the system. At the initial time $\tau_0$ the total state is
\begin{equation}
\begin{split}
		\ket{\Psi_0}=&  \int_2  \phi_0(x_1,x_2) \ket{x_1}_1\ket{x_2}_2\ket{\psi_0} dx_1 dx_2,
\end{split}
\end{equation} where the final ket (with no subscript) denotes a state of $S$, 
and the subscript on the integral sign indicates it is a double integral. 
This evolves to 
the following state immediately after the interaction with the first apparatus: 
\bqa
		\ket{\Psi_1}&=& \int_3  \phi_0(x_1,x_2)\ket{x_1+ X_1}_1\ket{x_2}_2 \nl{\times}\ket{X_1} \langle X_1\ket{\psi_0} dx_1dx_2dX_1.
\label{Psi1}
\eqa
Here $\ket{X_1}$ denote eigenstates of $\hat X_1\equiv\hat X(\tau_1)$, which we have taken to 
have a continuous spectrum for simplicity. 
Finally, after the second interaction, the total state is
\bqa
		\ket{\Psi_2}&=& \int_4  \phi_0(x_1,x_2) \ket{x_1+ X_1}_1\ket{x_2+ X_2}_2 \ket{X_2} \nl{\times}
 \bra{X_2}X_1\rangle\langle{X_1}\ket{\psi_0} dx_1dx_2dX_1 dX_2. \label{eq:total}
\eqa
From Eq. \eqref{eq:total}, the reduced state for the system at time $\tau_2^+$ is simply
\bqa
	\rho_2&=& \int_4 \phi_0^2\left(\frac{X_1-X_1'}{2},\frac{X_2-X_2'}{2}\right)	\ket{X_2}\ip{X_2}{X_1}\ip {X_1}{\psi}\nl{\times} \ip{\psi}{ X_1'} \ip{X_1'}{  X_2'} \bra{X_2'} dX_1 dX_2 dX_1'dX_2'.
\eqa

\section{All-at-once measurement at time $\tau_2^+$} 
It is convenient to use, rather than the observables $\hat z_n$ (\ref{defdisz}), the scaled observables
 \begin{equation}
	\begin{split}
		\hat y_1 &= \hat z_1/2 = \hat x_1+ a\hat x_2 \equiv \zeta_1(\hat x_1,\hat x_2),\\
		\hat y_2 &= \hat z_2/2 =\hat x_2+ a\hat x_1 \equiv \zeta_2(\hat x_1,\hat x_2).
	\end{split} \label{defdisy}
\end{equation}
A measurement of $\hat z_n$, or $\hat y_n$,  is described by the projector-density $\hat{\Pi}_n(y_n)$, defined by 
\begin{equation}\label{eq:projectors}
		\hat\Pi_n(y_n) =\int dx_1 \int  dx_2 \ \hat\pi_1(x_1)\otimes \hat\pi_2(x_2)  \delta(y_n-\zeta_n(x_1,x_2)), 
\end{equation} 
where $\hat\pi_n(x)=\ket{x}_n\bra{x}_n$. Note that, unlike $\hat\pi_n(x)$, $\hat \Pi_n(y)$ is  not a rank-one projector; it is in fact a rank-infinity projector. 
It satisfies $\int dy\hat \Pi_n(y)=1$ and $\hat \Pi_n(y)\hat \Pi_n(y')=\delta(y-y')\hat \Pi_n(y)$ (no sum over $n$ implied). It is obvious from the definition (\ref{defdisz}) that the two measurements commute.

Consider first the case where at time $\tau_2^+$ projective measurements of $\hat y_1$ and $\hat y_2$ are performed. This yields 
\bqa
\ket{\tilde\Psi_{2}(y_1,y_2)}&=& \hat \Pi_2(y_2)\hat \Pi_1(y_1) \ket{\Psi_2} \nn \\
&=& \ket{\frac{y_1-ay_2}{1-a^2}}_1\ket{\frac{y_2-ay_1}{1-a^2}}_2\ket{\tilde\psi_2(y_1,y_2)}, \nn\\
\label{eq:conditionstate}
\eqa
where the conditional system state $\ket{\tilde\psi_{2}(y_1,y_2)}$ is 
\bqa			
&&c\int_2 \exp[-(X_1-y_1)^2-(X_2-y_2)^2]
\nl{\times}\exp\ro{-2aX_1X_2-\frac{a^2(y_1^2+y_2^2)-2ay_1y_2}{1-a^2} }\ket{X_2}
\nl{\times}\bra{X_2}X_1\rangle\langle{X_1}\ket{\psi_0} dX_1 dX_2.
\label{eq:conditionsysstate}
\eqa
Obviously $S$ is no longer entangled with $\cu{A_1,A_2}$. This is as expected since the operators $\hat y_1$ and $\hat y_2$ are linearly independent, and jointly measuring these is equivalent to jointly measuring $\hat x_1$ and $\hat x_2$. That is, the measurement at time $\tau_2^+$ effects a rank-one projective measurement on the bath, disentangling it from the system. Moreover, it is easy to verify that, as expected, 
\beq
\frac{1}{1-a^2}\int_2\ket{\tilde\psi_{2}(y_1,y_2)}\bra{\tilde\psi_{2}(y_1,y_2)}  dy_1dy_2 = \rho_2.
\eeq
This establishes that \erf{eq:conditionsysstate} is indeed the discrete-time analogue of the solution of the NMSSE (\ref{NMSSE}) at the 
relevant time (here $\tau_2^+$).

\section{Monitoring (measurements at $\tau_1^+$ and $\tau_2^+$)} 
Now consider the case that \Dio\ claims is equivalent to the above, namely measuring  $\hat y_1$ at time $\tau_1^+$ and $\hat y_2$  at $\tau_2^+$. From \erf{Psi1}, the conditional total state at time $\tau_1^+$ is 
\bqa
	\label{Psi1alt}
		\ket{\tilde\Psi_{1}(y_1)}&=&\hat \Pi_1(y_1)\ket{\Psi_1}\\
		&=&\int e^{-\left(1-a^2\right) x^2}\ket{ {y_1} -ax}_1\ket{x}_2dx \ket{\tilde\psi_{1}(y_1)} , \nn
\eqa
where the conditional system state is
\beq 
\ket{\tilde\psi_{1}(y_1)} = c
\exp\left[- (y_1-\hat X_1)^2\right]\ket{\psi_0}.
\end{equation} So far we have a pure state for the system, as expected from \Dio's argument. However, at the very next step 
it breaks down. Because the measurement of the bath has disturbed it, we cannot use the state (\ref{eq:total}) to calculate the next conditioned state. Rather, we must calculate the effect of the interaction between $S$ and $A_2$ on state (\ref{Psi1alt}). The new entangled system-bath state at $\tau_2^+$ is 
\bqa
		\ket{\tilde\Psi_{2|1}(y_1)} &=&	\int_2 e^{-\left(1-a^2\right) x^2} \ket{y_1-ax}_1\ket{x+X_2}_2 dx  \nl{\times}\ket{X_2}\left\langle{X_2} \ket {\tilde\psi_{1}(y_1)}\right.  dX_2 \label{Psi21}
\eqa
Here the ${2|1}$ subscript indicates that the state is at time $\tau_2^+$ but the measurements it is conditioned upon was performed at time $\tau_1^+$. 

After the second measurement we have 
\beq
		\ket{\tilde\Psi_{2|1,2}(y_1,y_2)}=\hat \Pi_2(y_2)\ket{\tilde\Psi_{2|1}(y_1)},
\eeq
which evaluates to 
\bqa
&&c\int \ket{\frac{aX_2+y_1-ay_2}{1-a^2}}_1 \ket{\frac{y_2-a^2 X_2-ay_1 }{1-a^2}}_2\nl{\times}
\exp\left[-\frac{( X_2+ay_1-y_2)^2}{1-a^2}\right]
  \nl{\times}\ket{X_2}\bra{X_2} \exp[-(\hat X_1-y_1)^2]\ket{\psi_{0}}  dX_2 \label{Froghorn}
\eqa 
Note that this is an entangled state between $S$ and the bath --- it is not possible to define a pure conditional state
for the system. The reason is that, as noted above, the projector $\hat\Pi_2(y_2)$  is not rank-one, so there is no guarantee that it will disentangle the system from the bath. So the monitoring procedure \Dio\ describes cannot possibly correspond to the solution of the NMSEE (\ref{NMSSE}). Moreover, it is easy to verify that, as expected, 
\beq 
\int_2 \mathrm{Tr}_{12}\sq{\ket{\tilde\Psi_{2|1,2}(y_1,y_2)}\bra{\tilde\Psi_{2|1,2}(y_1,y_2)}} dy_1dy_2 \neq \rho_2.
\eeq
That is, the measurements described by \Dio\ disturb the evolution of the system so that it no longer obeys the original non-Markovian dynamics.

\section{Markovian limit} There is one case where \Dio's monitoring procedure does give a pure-state solution at all times which is identical to that which would be obtained by an all-at-once measurement at that time. This is case $a\rightarrow 0$, where $\hat y_n=\hat x_n$. That is to say, the intial bath state is unentangled, and the apparatuses are measured locally. In this Markovian limit we find  
\begin{equation}
	\ket{\tilde\Psi_{2}(y_1,y_2)}=  \ket{y_1}_1\ket{y_2}_2 \ket{\tilde\psi_{2}(y_1,y_2)},
\eeq
where the conditional state $\ket{\tilde\psi_{2}(y_1,y_2)}$ is given by
\beq
c\exp\left[- (\hat X_2-y_2)^2\right]\exp\left[- (\hat X_1-y_1)^2\right]\ket{\psi_0}.
\end{equation} 
This sequence of exponentials can obviously be continued indefinitely. 
The correspondence between the all-at-once measurement and \Dio's monitoring here is not surprising: in the Markovian limit the interpretation of a SSE in terms of continuous monitoring of the bath is well known.

To conclude, \Dio\ has introduced an elegant formulation of non-Markovian evolution using a local (Markovian) coupling to the bath but an initially non-local (entangled) bath state. In this formulation, it is simple to monitor the bath without affecting the future evolution of the system, because each apparatus only interacts with the system once. However, to make the conditioned system state pure, it is necessary to measure not only  the apparatuses which have already interacted with the system, but also some of those which are yet to interact. Measuring the latter necessarily introduces noise that will disturb the future evolution of the system, so that it will not reproduce the original non-Markovian evolution on average. We show by explicit calculation that the monitoring scheme suggested by \Dio\ does disturb the evolution in this manner, and moreover it  even fails to produce pure conditional system states. 

While it is certainly possible to derive a non-Markovian stochastic \sch\ equation, its solution can only be interpreted as a conditioned system state at some particular (but arbitrary) time $t$ \cite{GamWis02,GamWis03}. Connecting the solutions at different times creates the illusion of a ``quantum trajectory'', but is not part of standard quantum mechanics.  Rather, it is related to Bohmian mechanics and its generalizations \cite{GamWis04} which also allow one to derive discontinuous (jumplike) trajectories \cite{GamAskWis04}. Whether the jumplike non-Markovian trajectories recently introduced in Ref.~\cite{Piilo08} can be interpreted in a similar manner remains to be determined. But from the arguments in this Letter we know that 
non-Markovian pure-state trajectories cannot be interpreted as true quantum trajectories. 

{\em Acknowledgements:} HMW was supported by the Australian
Research Council grant FF0458313.
JMG was partially supported by MITACS and ORDCF.

\end{document}